\documentclass[9pt,twocolumn,twoside]{opticajnl}
\journal{opticajournal} %
\doi{}
\dates{}

\usepackage{xcolor}

\setboolean{shortarticle}{true}

\usepackage{siunitx}
\usepackage{lineno}
\usepackage{makecell}
\usepackage{graphicx}%
\usepackage{multirow}%
\usepackage{amsmath, amsfonts}%
\usepackage{mathrsfs}%
\usepackage{xcolor}%
\usepackage{textcomp}%
\usepackage{manyfoot}%
\usepackage{booktabs}%
\usepackage{algorithm}%
\usepackage{algorithmicx}%
\usepackage{algpseudocode}%
\usepackage{listings}%

\title{RTLViT: real-time lensless reconstruction with a lightweight vision transformer}

\author[1,*]{Leyla A. Kabuli}
\author[1]{Vasilisa Ponomarenko}
\author[1]{Laura Waller}

\affil[1]{Department of Electrical Engineering and Computer Sciences, University of California, Berkeley, CA 94720}

\affil[*]{lakabuli@berkeley.edu}

\begin{abstract}
Mask-based lensless imagers capture measurements using simple, compact hardware and recover images using a reconstruction algorithm. The reconstruction algorithm affects the image quality, inference speed, and computational requirements of the imaging system, often trading off image quality against computational efficiency. Advancing toward integrated lensless imagers, where encoding and reconstruction occur within a single device (e.g., a mobile phone), requires a fast, high-quality, and practical reconstruction method. We introduce the Real-Time Lensless Vision Transformer (RTLViT), a lightweight, purely data-driven reconstruction architecture for high-quality and real-time lensless reconstruction. With only 1.09 million learnable parameters, RTLViT provides higher-quality reconstructions than both real-time baselines and substantially larger attention-based architectures, including improvements of up to \SI{3.46}{\decibel} in peak signal-to-noise ratio over architectures with $15 \times$ more learnable parameters. We demonstrate that RTLViT maintains high-quality reconstructions across a wide range of training dataset sizes and two different mask designs (lenslets and a diffuser). Finally, we implement real-time reconstruction with RTLViT on laptops and smartphones, supporting future integrated lensless imagers for real-time applications.
\end{abstract}

\setboolean{displaycopyright}{false} %

\begin{document}

\maketitle

\begin{figure*}[hbtp]
\centering
    \includegraphics[width=\textwidth]{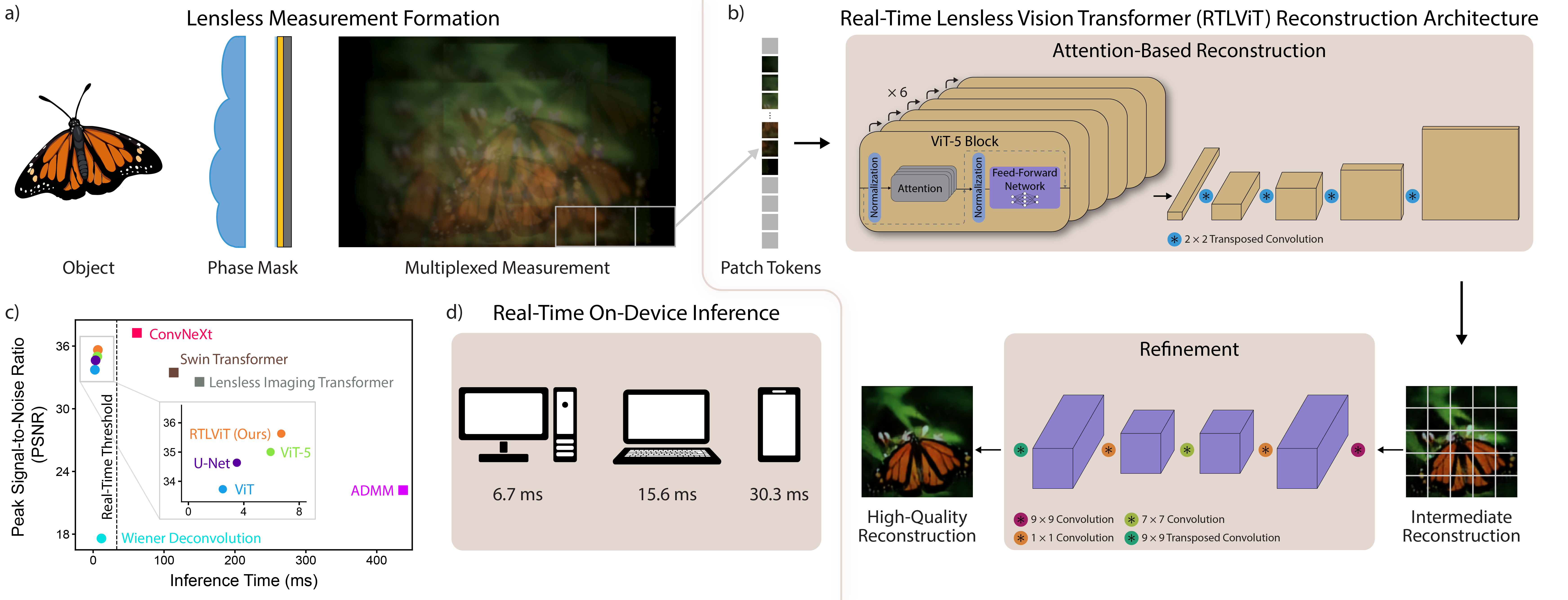}
    \caption{\textbf{Real-Time Lensless Vision Transformer (RTLViT) architecture for lensless reconstruction.} a) Lensless imagers produce multiplexed measurements.
    b) Measurements are decomposed into patch tokens and input to the attention-based reconstruction backbone, which consists of six Vision Transformer 5 (ViT-5) blocks and an upsampling decoder. The intermediate reconstruction is input to the convolutional refinement network, which merges patches and refines features, producing a high-quality reconstruction.
    c) Data-driven methods that achieve high-quality reconstructions (e.g., ConvNeXt~\cite{ConvRML}) for random multi-focal lenslet measurements are too slow for real-time inference. Fast physics-based methods (e.g., Wiener deconvolution) have low reconstruction quality, as quantified by the peak signal-to-noise ratio (PSNR). RTLViT provides the highest reconstruction quality among real-time data-driven methods.
    d) RTLViT achieves real-time inference on workstations, consumer laptops, and smartphones. 
    }
    \label{fig:intro}
\end{figure*}

Mask-based lensless imagers are compact, lightweight imaging systems that combine an optical encoder and computational reconstruction algorithm. Rather than producing an image directly, the optical encoder (e.g., a thin phase mask) produces a multiplexed sensor measurement (Fig.~\ref{fig:intro}a), which the reconstruction algorithm decodes into a reconstructed image. This multiplexing allows for higher-dimensional properties, including depth, to be captured in a single shot~\cite{DiffuserCam, PhlatCam}. 

The optical hardware design determines how much information is encoded into the measurement, and the reconstruction algorithm determines how effectively that information can be used to recover an image~\cite{LenslessInfoTheory}.
The algorithm choice also controls the \textit{inference time} for transforming a recorded measurement into a reconstruction, as well as practical requirements including algorithmic complexity, training data, memory usage, and necessary computing hardware. Real-time inference~\cite{jiminrealtimemicroscope}, defined here as $\leq$ \SI{33.3}{\milli\second} per reconstruction (30 frames per second (fps))~\cite{lenslessreview}, and lightweight models for integrated (on-device) deployment~\cite{DHMphone} are of growing interest for practical lensless imaging systems. %

Reconstruction methods have primarily been designed to trade off inference time and reconstruction quality (Fig.~\ref{fig:intro}c). The simplest approach, direct Wiener deconvolution, 
is real time and requires no training data, but provides poor-quality reconstructions. Iterative physics-based methods (e.g., alternating direction method of multipliers (ADMM)~\cite{ADMM}) require repeated optimization steps and are too slow for real-time inference. 
Hybrid approaches combine physics-based reconstruction with data-driven refinement to further improve reconstruction quality~\cite{MonakhovaLearning, mwdns}. Given sufficient training data, purely data-driven methods offer high reconstruction quality with an entirely learned approach~\cite{ConvRML, PanTransformer}.

In addition to inference time and reconstruction quality, practical implementation depends on calibration and computational requirements. 
Physics-based and hybrid methods require calibrated system point spread functions (PSFs) and explicit forward models, with high-quality methods often relying on additional spatially-varying PSF calibrations and corresponding multi-stage architectures~\cite{mwdns, jiminrealtimemicroscope}. Hybrid and purely data-driven methods have trended toward large networks with tens of millions of learnable parameters, high memory usage, large training datasets, and specialized computing hardware~\cite{MonakhovaLearning, ConvRML}. Balancing these practical requirements becomes particularly important in order to support real-time, on-device reconstruction and progress toward integrated lensless imaging systems.

Purely data-driven methods are promising for balancing inference time, reconstruction quality, and practical requirements. These methods learn a mapping from lensless measurements to ground truth images directly from training datasets without requiring calibrated PSFs or system modeling. However, improvements in reconstruction quality, including from convolutional~\cite{ConvRML} and attention-based~\cite{PanTransformer} architectures, have required increasingly large and slow models, along with extensive computational resources. 
Among real-time data-driven methods, previous lightweight Vision Transformer (ViT) architectures have poor reconstruction quality~\cite{ConvRML}, while baseline architectures such as U-Nets are substantially larger~\cite{UNet, MonakhovaLearning}.
There remains a need for a high-quality, lightweight data-driven architecture suitable for real-time and on-device deployment.

In this work, we introduce the Real-Time Lensless Vision Transformer (RTLViT), a purely data-driven, attention-based reconstruction architecture designed to simultaneously achieve high-quality reconstruction, real-time inference, and minimal computational requirements. With only 1.09 million learnable parameters, our lightweight RTLViT outperforms existing real-time architectures and rivals substantially larger and slower data-driven architectures in reconstruction quality across training dataset sizes for both lenslet- and diffuser-based systems. We further demonstrate real-time reconstruction on a consumer laptop and a smartphone, establishing RTLViT's suitability for low-compute, on-device lensless reconstruction.

Lensless measurement formation follows the forward model $y = \mathcal{N}(A x).$ The measurement $y$ corresponds to an object $x$ that is encoded by a multiplexing matrix $A$, defined by the system PSFs and mapped through measurement noise $\mathcal{N}(\cdot)$.
Instead of incorporating this forward model or calibrated system PSFs for reconstruction, purely data-driven methods such as RTLViT learn a mapping $f_\theta(\cdot)$ from measurements to objects directly from a dataset of $K$ paired lensless measurements. This corresponds to optimizing a set of parameters $\theta$ such that
\begin{equation}
\theta^\ast = 
\underset{\theta}{\arg\min} \frac{1}{K} \sum_{k=1}^{K} || x_k - f_{\theta}(y_k)||_2^2,
    \label{eq:lossfunction}
\end{equation}
where $f_{\theta}$ is a network that takes as input a measurement $y_k$ and outputs an object estimate $f_\theta(y_k)$. Network parameters are optimized based on reconstruction similarity to each object $x_k$, which we quantify using a mean squared error (MSE) loss. 

RTLViT combines an attention-based reconstruction backbone with a convolutional neural network (CNN) refiner (Fig.~\ref{fig:intro}b). Attention is used to process long-range dependencies in lensless measurements, which are introduced by multiplexing PSFs~\cite{PanTransformer}. The attention-based backbone produces an intermediate reconstruction, which is processed by the CNN to recover fine features and correct blocking artifacts.

\begin{figure}[t]
\centering
\includegraphics[width=0.85\linewidth]{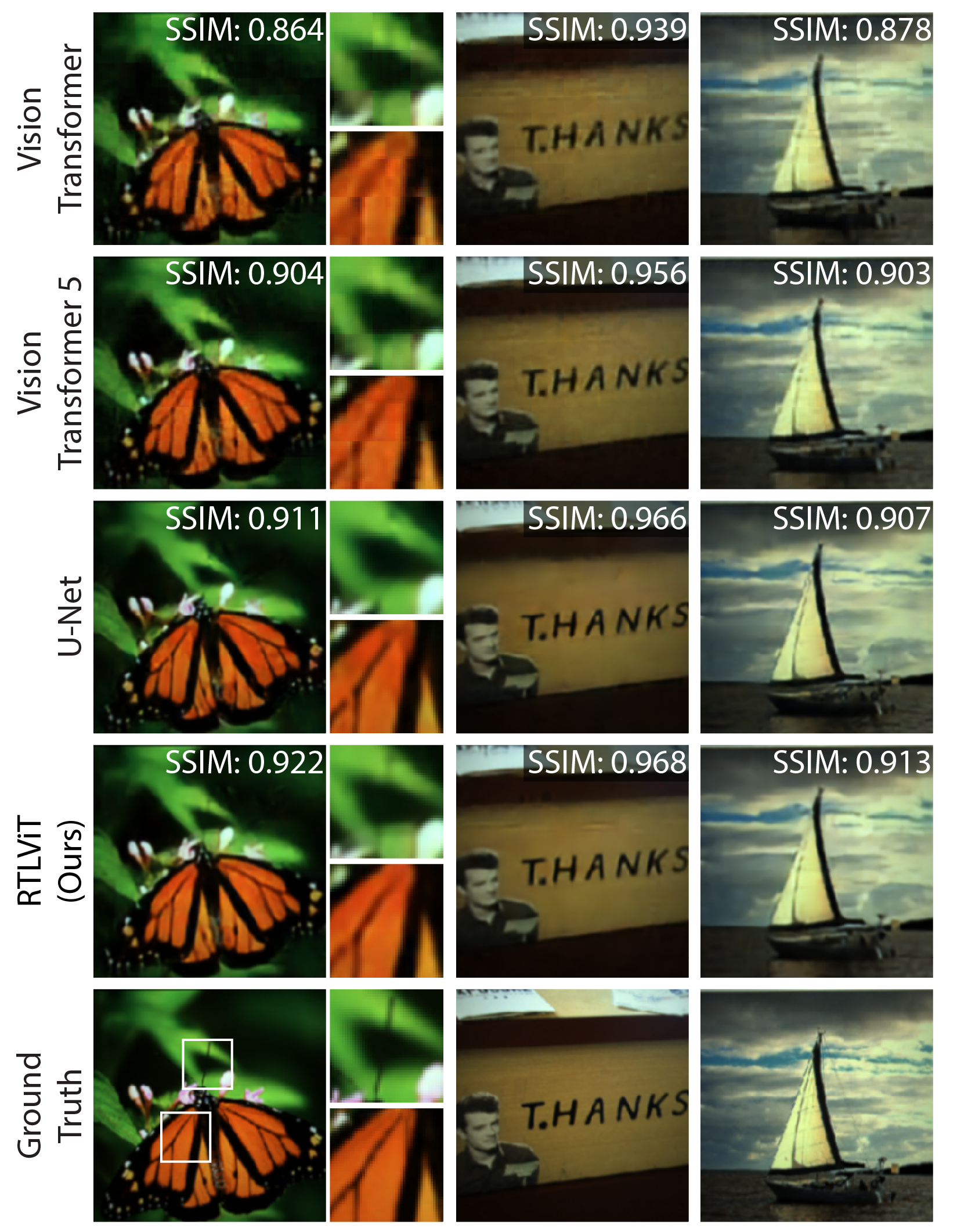}
\caption{\textbf{Random multi-focal lenslet reconstruction examples.} The Vision Transformer 5 (ViT-5) component improves reconstruction quality over the ViT, as quantified by the structural similarity index measure (SSIM). The complete RTLViT eliminates blocking artifacts and better recovers fine features than the U-Net, as highlighted in the butterfly insets. Corresponding reconstruction metrics are reported in Table~\ref{tab:rml_comparison}.
}
\label{fig:architecture_components}
\end{figure}

The backbone is a reconstruction-specific, lightweight adaptation of a Vision Transformer 5 (ViT-5)~\cite{vit5}. A ViT represents an image as a sequence of patch tokens and processes them using self-attention. The ViT-5 modernizes this architecture with updated components including positional encoding and normalization (Supplement Table~S1). 
Our backbone is smaller than a typical ViT-5, using six transformer blocks with 128-dimensional token embeddings, four attention heads, and a feed-forward network with a 256-dimensional hidden layer. Each transformer block follows the ViT-5 structure: inputs are normalized before self-attention and again before the feed-forward network, with residual connections around both operations. 
We extend the ViT-5 rotary positional embeddings (RoPE)~\cite{vit5} to support rectangular lensless measurements and cache fixed RoPE tables for a $2\times$ speedup (Supplement Sec.~1).
For lensless image reconstruction, we find that this ViT-5-style architecture improves reconstructions compared to a parameter-matched ViT~\cite{ConvRML} (Fig.~\ref{fig:architecture_components}). 

Each measurement is divided into non-overlapping $15 \times 15$ pixel patches and five learnable tokens are added during ViT-5 encoding~\cite{vit5}. After passing through the six ViT-5 blocks, the learnable tokens are removed and the patch tokens are reshaped into a low-resolution spatial feature map. Our compact decoder then applies four $2 \times 2$ transposed convolution layers with stride 2, batch normalization, and ReLU activation, followed by bilinear interpolation to produce the intermediate reconstruction. 

Using non-overlapping measurement patches in our backbone minimizes architectural complexity, but can introduce discontinuities at patch boundaries, or blocking artifacts, as shown in the butterfly insets for the ViT and ViT-5 in Fig.~\ref{fig:architecture_components}. Transformer architectures with overlapping patches or sliding windows can mitigate these blocking artifacts, but previous implementations for lensless imaging resulted in substantially larger and slower models without improvements in reconstruction quality~\cite{ConvRML}. 

To remove blocking artifacts and further improve reconstruction quality, we instead incorporate a refinement network. We select a Fast AR-CNN~\cite{fastarcnn}, a fast and lightweight network designed to remove blocking artifacts from image compression, which are qualitatively similar to the blocking artifacts present in our reconstructions. Fast AR-CNN applies a $9 \times 9$ convolution with stride 2, a $1 \times 1$ channel reduction, a $7 \times 7$ convolution, a $1 \times 1$ channel expansion, and a $9 \times 9$ transposed convolution with stride 2, with parametric rectified linear unit (PReLU) activations between layers. Within RTLViT, this refinement network suppresses blocking artifacts and faithfully recovers fine image features (Fig.~\ref{fig:architecture_components}), producing the final high-quality reconstruction.

RTLViT has 1,091,290 learnable parameters, of which 1,005,715 are from the ViT-5-based backbone and 85,575 are from the Fast AR-CNN. The two components are trained jointly for 70 epochs using learning rates of $5 \times 10^{-3}$ and $1 \times 10^{-3}$, respectively. Model training uses the AdamW optimizer with weight decay $10^{-3}$, linear warmup, cosine annealing, MSE loss~(\eqref{eq:lossfunction}), a batch size of 6, and one NVIDIA A6000 GPU. For each model, the checkpoint with the lowest validation loss is reported.

We evaluate architectures on natural images from the Parallel Lensless Dataset (PLD)~\cite{ConvRML}, using 45,000 measurements for training, 4,000 for validation, and 1,000 for testing unless otherwise specified. Measurements are downsampled to $300 \times 480$ pixels and normalized to $[0, 1]$. Metrics are evaluated on $214 \times 214$ pixel image regions and averaged over the test set.

Among existing purely data-driven methods suitable for real-time reconstruction, we select the ViT~\cite{ConvRML} with 1,006,867 learnable parameters as the lightweight baseline and the U-Net~\cite{MonakhovaLearning, UNet} with 11,772,307 learnable parameters as a representative standard reconstruction architecture.  We follow the same training procedure as RTLViT for the ViT and U-Net with a learning rate of $5 \times 10^{-4}$, along with 35 training epochs for the U-Net.

We first evaluate these architectures on random multi-focal lenslet (RML) measurements from the PLD, representing a low-multiplexing system (sparse PSF) with relatively high-quality measurements. RTLViT achieves the highest test set peak signal-to-noise ratio (PSNR) and structural similarity index measure (SSIM) among evaluated models (Table~\ref{tab:rml_comparison}), despite having $10.8 \times$ fewer learnable parameters than the U-Net. Although the U-Net produces smoother reconstructions than the ViT-5 component and achieves lower learned perceptual image patch similarity (LPIPS) than RTLViT, RTLViT faithfully recovers fine details, such as the butterfly antenna, that the U-Net does not (Fig.~\ref{fig:architecture_components}).  

\begin{table}[!b]
\centering
\caption{Reconstruction quality evaluated on low-multiplexing random multi-focal lenslet measurements.}
\label{tab:rml_comparison}
\begin{tabular}{lcccc}
\toprule
\textbf{Model}
& \textbf{PSNR $\uparrow$}
& \textbf{SSIM $\uparrow$}
& \textbf{LPIPS $\downarrow$}
\\ 
\midrule
ViT        
& 33.732 & 0.886 & 0.147 \\
U-Net   
& 34.636 & 0.915  & \textbf{0.0919} \\
RTLViT (Ours)        
&  \textbf{35.630} & \textbf{0.921}  & 0.107 \\
\bottomrule
\end{tabular}
\end{table} 

\begin{table}[hbtp]
\centering
\caption{Reconstruction quality evaluated on high-multiplexing diffuser measurements.}
\label{tab:diffuser_comparison}
\begin{tabular}{lcccc}
\toprule
\textbf{Model}
& \textbf{PSNR $\uparrow$}
& \textbf{SSIM $\uparrow$}
& \textbf{LPIPS $\downarrow$}
\\ 
\midrule
ViT        
& 27.186 & 0.733 & 0.333 \\
U-Net   
& 27.646 & 0.754  & 0.299 \\
RTLViT (Ours)        
&  \textbf{28.712} & \textbf{0.783}  & \textbf{0.282} \\
\bottomrule
\end{tabular}
\end{table}

We then evaluate the same architectures trained on diffuser measurements from the PLD, representing a high-multiplexing system (less sparse PSF) with reduced measurement contrast. RTLViT achieves the highest reconstruction quality across quantitative metrics (Table~\ref{tab:diffuser_comparison}), consistent with reconstruction examples shown in Supplement Fig.~S1.
The ViT-5 alone has higher SSIM than the U-Net, highlighting the benefits of attention for capturing the long-range dependencies introduced by multiplexed measurements.

\begin{figure}[ht]
\centering
\includegraphics[width=0.75\linewidth]{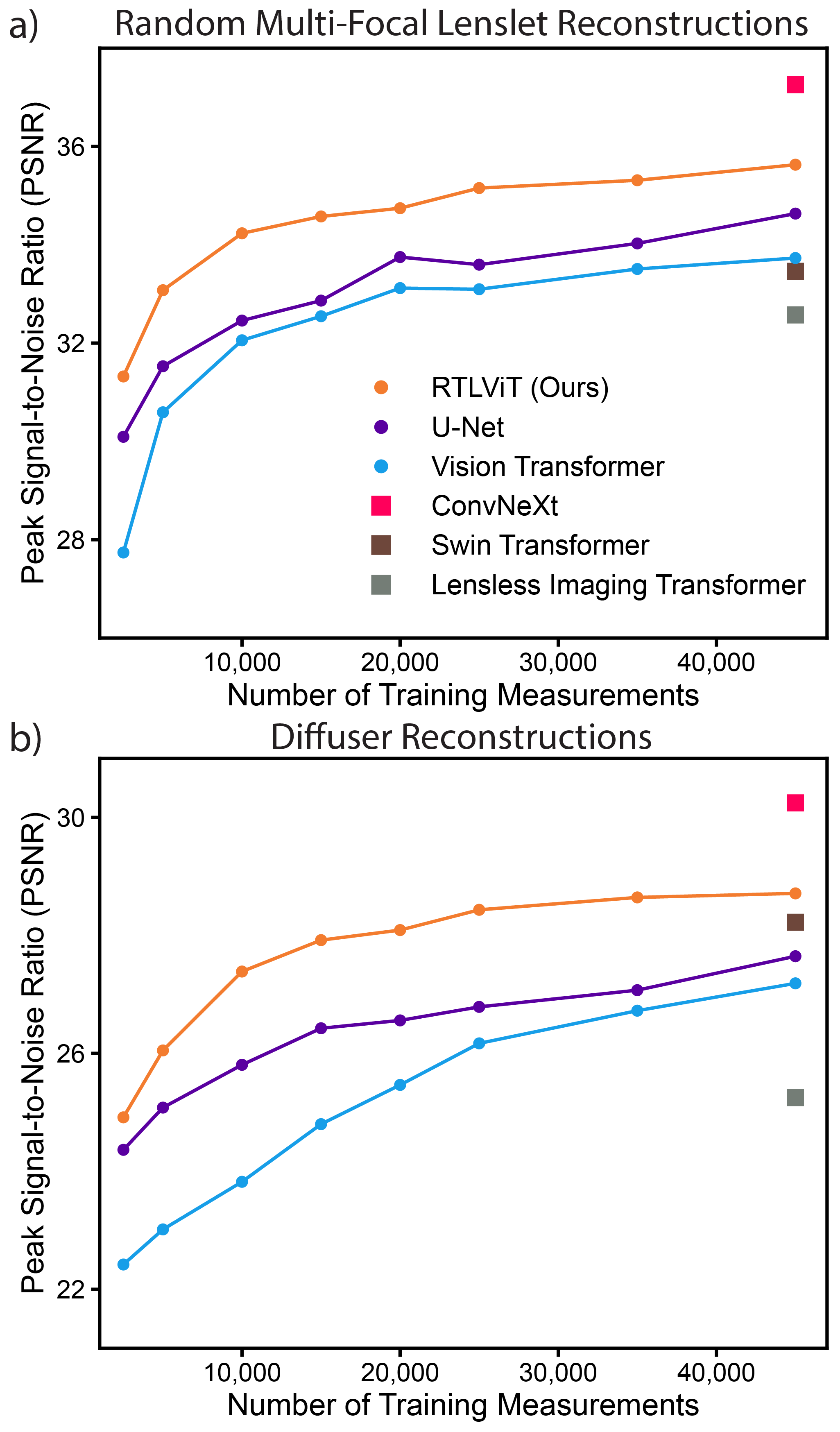}
\caption{\textbf{Reconstruction quality across training dataset sizes for two different lensless imaging systems.} RTLViT has higher peak signal-to-noise ratio (PSNR) than the U-Net and the Vision Transformer across all dataset sizes for both the a) random multi-focal lenslet design and b) the diffuser design, and is closest to ConvNeXt. Square markers denote architectures that are not suitable for real-time reconstruction.}
\label{fig:dataset_sweep}
\end{figure}

Next, we assess reconstruction quality across training dataset sizes for both RML and diffuser systems. Keeping validation and testing datasets fixed, we train individual ViT, U-Net, and RTLViT models for training dataset sizes ranging from 2,500 to 45,000 measurements.
RTLViT maintains the highest PSNR across all dataset sizes and both imaging systems (Fig.~\ref{fig:dataset_sweep}).

For comprehensive reconstruction quality assessment, Fig.~\ref{fig:dataset_sweep} includes comparisons with larger data-driven architectures previously evaluated on the PLD: the Lensless Imaging Transformer (LIT)~\cite{PanTransformer}, Swin Transformer~\cite{ConvRML}, and ConvNeXt~\cite{ConvRML}. These architectures are high-quality references and are not suitable for real-time inference (Fig.~\ref{fig:intro}c) or lightweight, on-device deployment. Starting at only 10,000 training measurements for the RML and 25,000 for the diffuser system, RTLViT already has higher PSNR than both the LIT and Swin Transformer, which were each trained using 45,000 measurements. At the full dataset size, RTLViT improves PSNR by \SI{3.46}{\decibel} over the LIT for diffuser reconstructions while using $15 \times$ fewer learnable parameters.
Among the evaluated architectures, only ConvNeXt achieves higher PSNR. RTLViT remains within \SI{1.63}{\decibel} PSNR for RML and \SI{1.53}{\decibel} for diffuser measurements, despite using $33 \times$ fewer learnable parameters than ConvNeXt. Similar trends are observed for SSIM and LPIPS (Supplement Fig.~S2). 

To establish RTLViT as a lightweight architecture, we characterize its computational requirements. RTLViT has a model size of \SI{5.36}{\mega\byte}, uses 8.68 giga floating-point operations (GFLOPs) per forward pass, and uses \SI{1.8}{\giga\byte} of GPU memory during training (Supplement Tables~S2 and~S5). These requirements remain lightweight, significantly lower than those of the U-Net and only modest increases relative to the ViT and ViT-5.

Finally, we demonstrate real-time, on-device deployment of RTLViT on a workstation (NVIDIA A6000 GPU), a consumer laptop (MacBook Pro M1), and a smartphone (iPhone 17 Pro). We evaluate the RTLViT model trained on 45,000 RML measurements and report average inference time over 100 reconstructions with GPU synchronization following 10 warm-up runs. For smartphone deployment, we convert the trained RTLViT model from PyTorch to Apple Core ML and implement inference in a simple Swift-based iPhone application. RTLViT maintains real-time inference across all three computing platforms (Fig.~\ref{fig:intro}d). Workstation inference time can be reduced to \SI{1.8}{\milli\second} with PyTorch compilation (Supplement Table~S4). Notably, our \SI{30.3}{\milli\second} (33 fps) reconstructions on the smartphone exceed the real-time reconstruction threshold of 30 fps.
To our knowledge, this is the first demonstration of mask-based lensless image reconstruction on a smartphone, and it operates in real time.

In summary, we present a lightweight, purely data-driven reconstruction architecture that combines high-quality reconstruction, real-time inference, and low computational requirements. Across both lenslet- and diffuser-based imaging systems, RTLViT achieves the highest reconstruction quality among real-time architectures and is competitive with much larger architectures for high-quality reconstruction. RTLViT reconstructs measurements in real time on a consumer laptop and a smartphone. These capabilities can be used for applications requiring real-time visualization, including microscopy~\cite{jiminrealtimemicroscope} and compact surgical and endoscopic imaging. 
RTLViT does not require a forward model or system calibration, but it does require training datasets and retraining following system changes. Our lightweight architecture may inform future real-time hybrid reconstruction architectures~\cite{jiminrealtimemicroscope, mwdns}, extensions to 3D reconstruction, and general deconvolution problems. %

\begin{backmatter}
\bmsection{Funding} U.S.A.F. Office of Scientific Research (FA955-22-1-0521).

\bmsection{Acknowledgment} The authors thank Mengyue Geng and April Chen for support. The authors used OpenAI Codex only to assist with development of the smartphone application. L.W. is a Chan Zuckerberg Biohub SF investigator. L.W. is on appointment as a Miller Research Professor in the Miller Institute for Basic Research in Science. 

\bmsection{Disclosures} 
The authors declare no conflicts of interest.
\smallskip

\bmsection{Data availability} Source code will be made publicly available~\cite{RTLViT_repo}.

\bmsection{Supplemental document}
See Supplementary Material for supporting content. 
\end{backmatter}

\bibliography{sample}

\bibliographyfullrefs{sample}

\clearpage

\renewcommand{\thesection}{S\arabic{section}}
\renewcommand{\thetable}{S\arabic{table}}
\renewcommand{\thefigure}{S\arabic{figure}}
\renewcommand\theequation{S\arabic{equation}}
\setcounter{figure}{0}
\setcounter{table}{0}
\setcounter{section}{0}
\setcounter{equation}{0}
\onecolumn

\noindent
\textbf{\huge RTLViT: real-time lensless reconstruction with a lightweight vision transformer \\
Supplementary Material}

\vspace{0.1cm}

\noindent 
\textbf{\large Leyla~A.~Kabuli$^{1, \ast}$, Vasilisa~Ponomarenko$^{1}$, and Laura~Waller$^{1}$}

\noindent
\small$^1$Department of Electrical Engineering and Computer Sciences, University of California, Berkeley, CA 94720\\
\small$^\ast$Corresponding author. Email: lakabuli@berkeley.edu 
\\

\noindent 
This supplementary material contains additional results and implementation details.

\section{ViT-5 Architecture}
\label{sec:vit5}

We summarize the updates to the Vision Transformer (ViT) architecture provided by ViT-5~\cite{vit5}. ViT-5 replaces LayerNorm with root mean square normalization (RMSNorm), removes biases from query, key, and value representations, and applies RMSNorm to queries and keys within each attention head (QK normalization). Each ViT-5 transformer block includes two learnable LayerScale vectors, which are applied to the outputs of the attention and feed-forward components of the block. 

ViT-5 uses both absolute positional embeddings and 2D rotary positional embeddings (RoPE) for patch tokens. Five learnable tokens, one global and four register, are included for self-attention to provide a flexible representation space. A separate register RoPE is applied to the four register tokens. We use a theta value of 10,000 for patch RoPE and a theta value of 100 for register RoPE, following the ViT-5 implementation~\cite{vit5}. The theta value controls the frequencies used for rotation, with a smaller theta value corresponding to higher frequencies. The use of separate theta values is intended to reduce positional correlations between patch and register tokens~\cite{vit5}.

The original 2D RoPE implementation for ViT-5 assumes a square patch grid. We extend this to rectangular lensless measurements by specifying separate height and width and corresponding frequencies for the rectangular patch grid. We also precompute and cache the RoPE frequency tables instead of rebuilding them in each forward pass. This update reduces ViT-5 inference time by $2\times$, from \SI{11.8}{\milli\second} without caching to \SI{5.9}{\milli\second}.

We evaluate the effect of each updated ViT-5 component using random multi-focal lenslet (RML) measurements from the Parallel Lensless Dataset (PLD)~\cite{ConvRML} in Table~\ref{tab:vit5_ablation}. All evaluations use models trained on 45,000 measurements following the training procedure outlined in the main text. For each evaluation, one component is removed or modified compared to the complete ViT-5 architecture. We test removing LayerScale, QK normalization, patch RoPE, and learnable register tokens, as well as increasing the theta value for register RoPE to 10,000, equal to that of patch RoPE. Each modification to the architecture reduces reconstruction quality, resulting in lower peak signal-to-noise ratio (PSNR) and structural similarity index measure (SSIM), and higher learned perceptual image patch similarity (LPIPS). Removing LayerScale or QK normalization results in the largest decrease in reconstruction quality, while increasing the register RoPE theta value to 10,000 has the least effect on reconstruction quality. The complete ViT-5 architecture results in the best reconstruction quality, indicating that each updated ViT-5 component has at least a modest benefit for lensless image reconstruction.

\begin{table}[ht]
\centering
\caption{Effect of Vision Transformer 5 (ViT-5) architecture components on lensless reconstruction}
\label{tab:vit5_ablation}
\begin{tabular}{lcccc}
\toprule
\textbf{Model Configuration}
& \textbf{PSNR $\uparrow$}
& \textbf{SSIM $\uparrow$}
& \textbf{LPIPS $\downarrow$}
\\ 
\midrule
Complete ViT-5 & \textbf{35.005} & \textbf{0.910} & \textbf{0.112} \\ 
Remove LayerScale & 34.358 & 0.898 & 0.124 \\ 
Remove QK normalization & 34.503 & 0.900 & 0.125 \\
Remove RoPE & 34.675 & 0.904 & 0.121 \\
Remove register tokens & 34.695 & 0.904 & 0.116 \\
Equal register and patch theta & 34.848 & 0.907 & 0.116 \\
\bottomrule
\end{tabular}
\end{table}

\section{Training details}

As described in the main text, reported models correspond to the model checkpoint with the best validation loss. 
This corresponds to epoch 70 for the ViT, epoch 34 for the U-Net, and epoch 68 for the RTLViT trained on 45,000 RML measurements from the PLD. For models trained on 45,000 diffuser measurements from the PLD, this corresponds to epoch 70 for the ViT, epoch 33 for the U-Net, and epoch 69 for the RTLViT. We use pretrained ConvNeXt~\cite{ConvRML} models for both RML and diffuser baselines. For the Swin Transformer~\cite{ConvRML} and the Lensless Imaging Transformer~\cite{PanTransformer} in Fig.~3 in the main text, we use pretrained models for RML measurements~\cite{ConvRML}. As corresponding pretrained models for diffuser measurements are not available, we train new models for diffuser measurements following the same training procedure~\cite{ConvRML}. Reported reconstruction metrics on diffuser measurements correspond to epoch 35 for the Swin Transformer and step 185,000 ($\approx$ epoch 25) for the Lensless Imaging Transformer. 

Training time, training epochs, and peak GPU memory usage for real-time compatible architectures are reported in Table~\ref{tab:pld_train_time}. RTLViT training time is comparable to that of the ViT-5, and all attention-based architectures maintain lower peak GPU memory than the U-Net. 

For the physics-based reconstructions of RML measurements in Fig.~1c in the main text, we use the RML point spread function (PSF) provided in the PLD~\cite{ConvRML}. We use CPU implementations of Wiener deconvolution with a regularization parameter of $10^{-2}$ and ADMM with 10 iterations and parameters $\mu_1 = 1 \times 10^{-4}$, $\mu_2 = 1 \times 10^{-4}$, $\mu_3 = 1 \times 10^{-4}$, $\tau = 2 \times 10^{-3}$. Reconstruction metrics are averaged over the test set. Inference times are calculated following the procedure described in the main text.

\begin{table*}[h]
\centering
\caption{Training time, epochs, and GPU memory for real-time compatible models using 45,000 training and 4,000 validation measurements from the Parallel Lensless Dataset}
\label{tab:pld_train_time}
\begin{tabular}{lccc}
\toprule
\textbf{Model}
& \textbf{Training time (h)} 
& \textbf{Epochs}
& \textbf{GPU memory (GB)} \\
\midrule

ViT & 5.5 & 70 & 1.0\\
ViT-5 & 9 & 70 & 1.4 \\
U-Net & 5 & 35 & 3.0  \\
RTLViT (Ours) & 9.5 & 70 & 1.8 \\

\bottomrule
\end{tabular}
\end{table*}

\section{Additional Reconstruction Results}
\begin{figure}[h]
\centering
\includegraphics[width=0.65\linewidth]{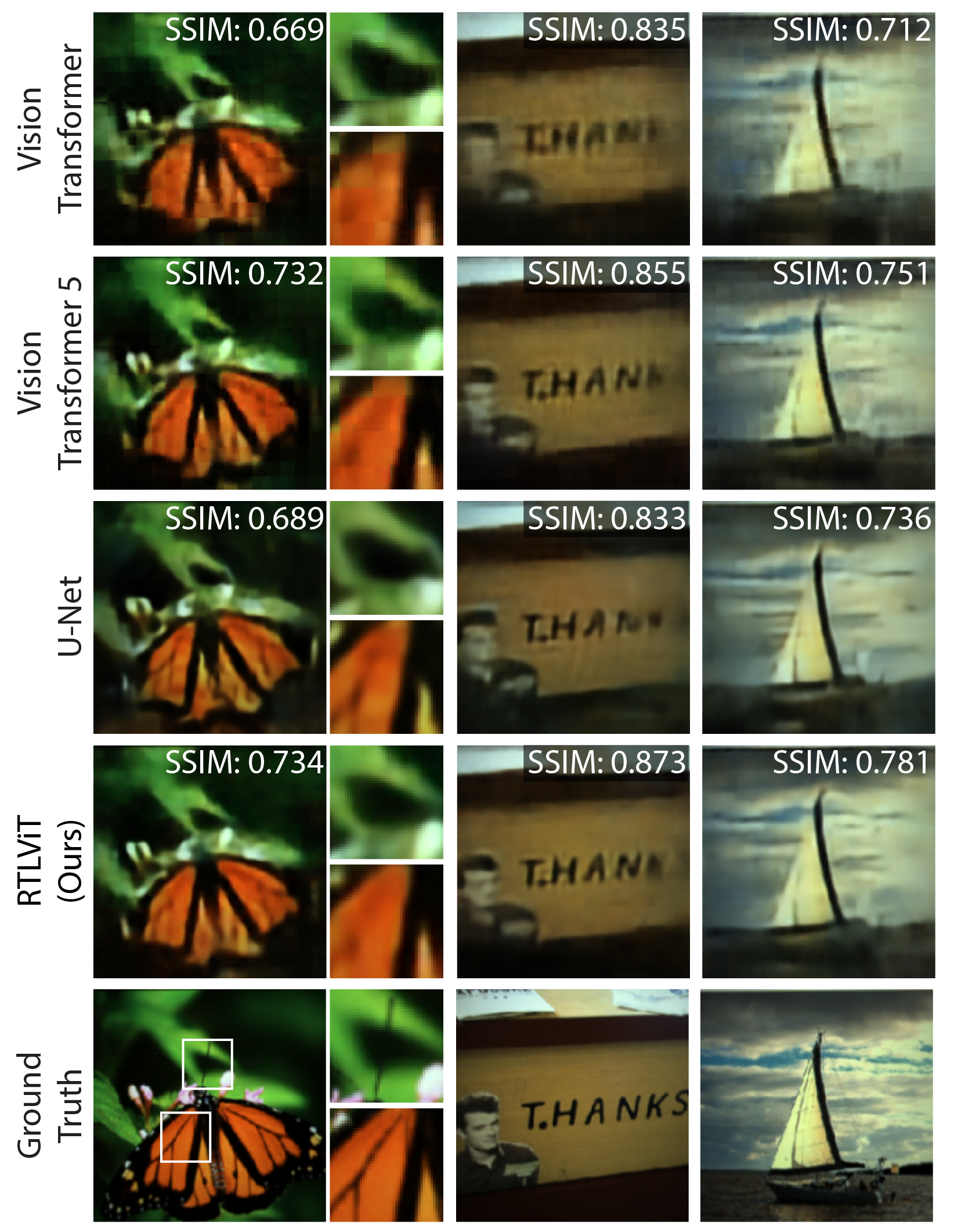}
\caption{\textbf{Diffuser reconstruction examples.} The Vision Transformer 5 (ViT-5) component improves reconstruction quality over the ViT and the U-Net, as quantified by the structural similarity index measure (SSIM). RTLViT eliminates blocking artifacts and achieves the highest-quality reconstructions among evaluated models.
}
\label{fig:supp_architecture_components}
\end{figure}

We include reconstructions for high-multiplexing diffuser measurements from the PLD in Fig.~\ref{fig:supp_architecture_components}, complementing the reconstruction quality metrics in Table~2 in the main text. Unlike the RML in Fig.~2 in the main text, ViT-5 alone has higher SSIM than the U-Net for diffuser reconstructions. RTLViT has the highest-quality reconstructions, consistent with quantitative evaluations. 

\begin{figure}[]
\centering
\includegraphics[width=0.7\linewidth]{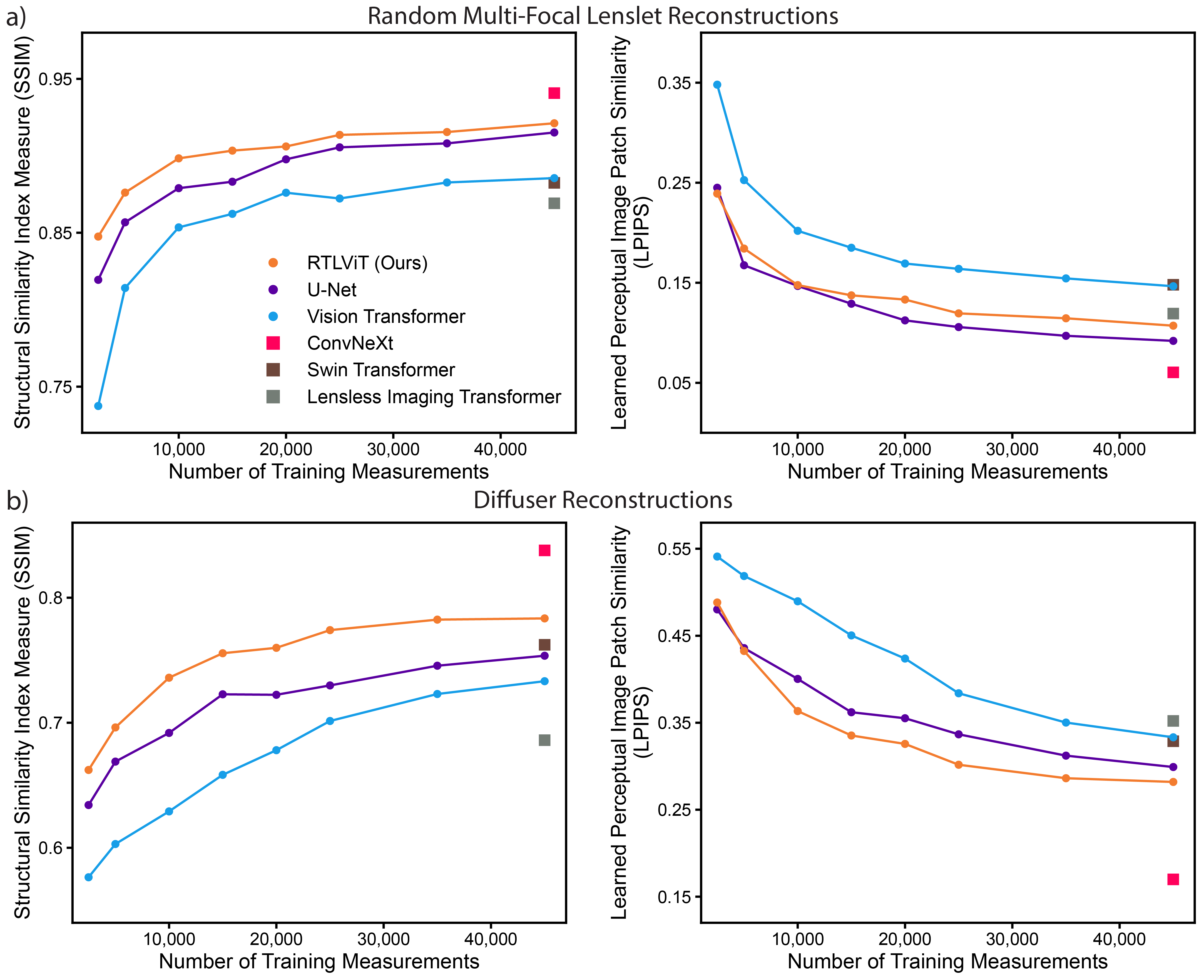}
\caption{\textbf{Reconstruction quality across training dataset sizes for two different lensless imaging systems.} 
RTLViT has higher structural similarity index measure (SSIM) than the U-Net and the Vision Transformer (ViT) across all dataset sizes for both the a) random multi-focal lenslet and b) diffuser designs, and is closest to ConvNeXt among evaluated models. RTLViT also has lower learned perceptual image patch similarity (LPIPS) than the U-Net and the ViT for diffuser measurements.}
\label{fig:supp_dataset_sweep}
\end{figure}

Figure~\ref{fig:supp_dataset_sweep} reports SSIM and LPIPS for the training dataset size evaluations from Fig.~3 in the main text. RTLViT has higher SSIM than the ViT and the U-Net across all evaluated training dataset sizes for both imaging systems. RTLViT has lower LPIPS than the U-Net for diffuser measurements, whereas the U-Net has slightly lower LPIPS than the RTLViT for RML measurements.

We additionally evaluate RTLViT on the DiffuserCam Lensless Mirflickr Dataset (DLMD)~\cite{MonakhovaLearning}. RTLViT achieves the best PSNR, SSIM, and LPIPS among evaluated real-time compatible models (Table~\ref{tab:dlmd_comparison}), consistent with the PLD diffuser results in the main text. 

For the DLMD, models are trained following the training procedure described in the main text, with 24,000 measurements used for training and 1,000 measurements used for validation and testing. Reported model checkpoints are selected based on the lowest validation loss, corresponding to epoch 67 for the ViT, epoch 69 for the RTLViT, and epoch 30 for the U-Net. With $270 \times 480$ pixel measurements from the DLMD, the ViT parameter count is reduced to 998,675, while the ViT-5 component and complete RTLViT parameter counts are reduced to 997,523 and 1,083,098, respectively. The U-Net and Fast AR-CNN component parameter counts remain unchanged.

\begin{table}[hbtp]
\centering
\caption{Reconstruction quality evaluated on the DiffuserCam Lensless Mirflickr Dataset}
\label{tab:dlmd_comparison}
\begin{tabular}{lcccc}
\toprule
\textbf{Model}
& \textbf{PSNR $\uparrow$}
& \textbf{SSIM $\uparrow$}
& \textbf{LPIPS $\downarrow$}
\\ 
\midrule
ViT%
& 21.493 & 0.635 & 0.415 \\ %
U-Net   
& 21.546 & 0.662  & 0.414 \\ %
RTLViT (Ours)        
&  \textbf{23.168} & \textbf{0.694}  & \textbf{0.382} \\ %

\bottomrule
\end{tabular}
\end{table}

\section{Computational Requirements}

\begin{figure}[]
\centering
\includegraphics[width=0.25\textwidth]{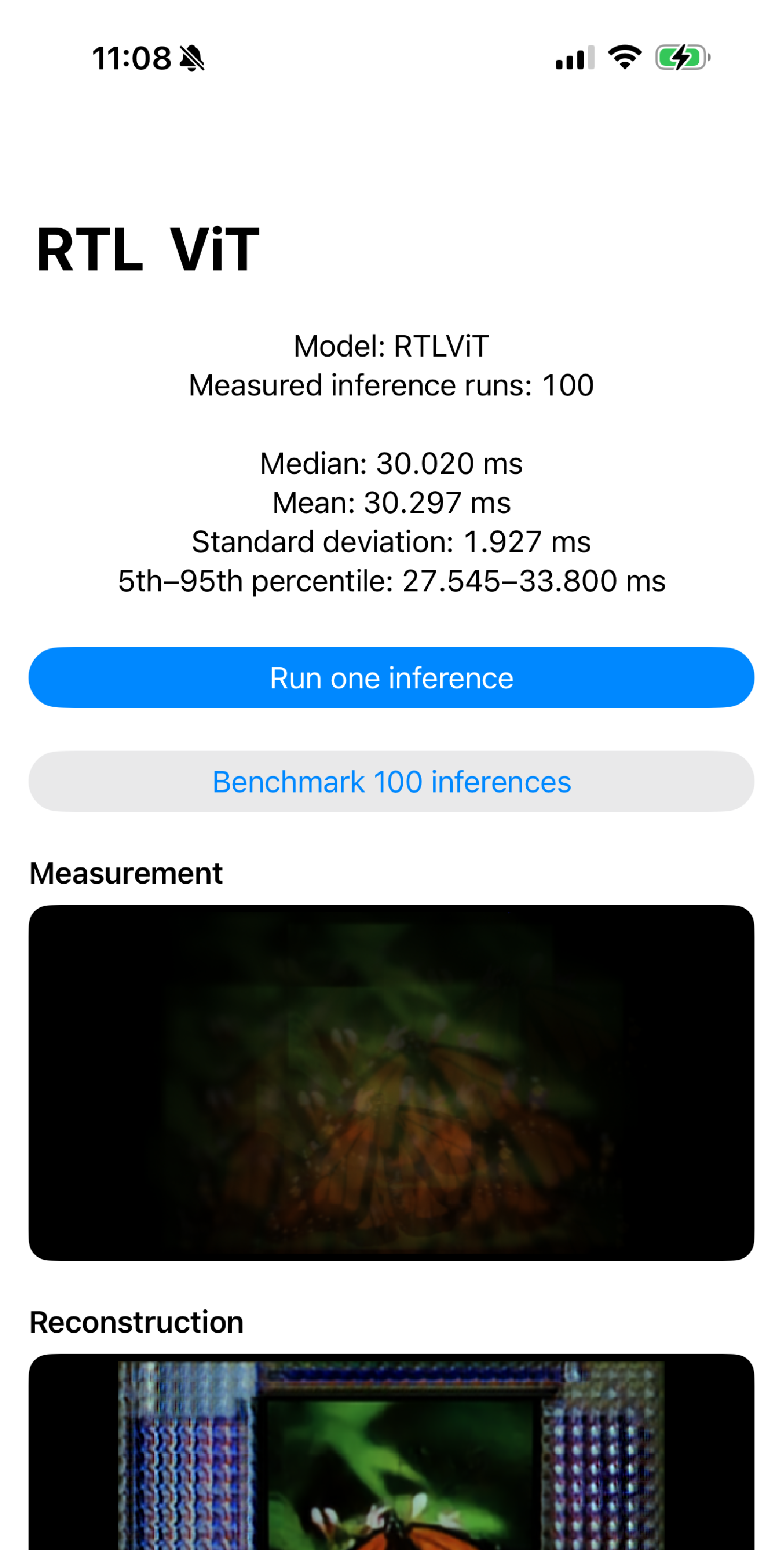}
\caption{\textbf{Smartphone application for lensless reconstruction.} The graphical user interface displays a measurement and its reconstruction, and supports evaluating inference time.}
\label{fig:supp_phone}
\end{figure}

In Table~\ref{tab:comparison_inference}, we extend the inference time comparison in Fig.~1d in the main text to additional reconstruction architectures. We evaluate a workstation with an NVIDIA A6000 GPU, a 16-core CPU, and \SI{128}{\giga\byte} RAM, a MacBook Pro with an Apple M1 Pro chip, a 10-core CPU, and \SI{32}{\giga\byte} RAM, and an iPhone 17 Pro. Inference times are calculated following the procedure described in the main text. A screenshot of the graphical user interface for the smartphone application is shown in Fig.~\ref{fig:supp_phone}.

As noted in the main text, RTLViT maintains real-time inference across the three computing platforms. On the workstation, the ViT-5 is slower than the ViT, but adding the Fast AR-CNN refinement network for RTLViT increases inference time by only \SI{0.8}{\milli\second}, from \SI{5.9}{\milli\second} to \SI{6.7}{\milli\second}. RTLViT is slightly faster than the U-Net on the laptop, but slower on the smartphone. ConvNeXt, included as a high-quality reference, requires \SI{61.7}{\milli\second} on the workstation and \SI{535}{\milli\second} on the laptop, greatly exceeding the threshold for real-time reconstruction. Therefore, ConvNeXt is not included for smartphone evaluation. We also do not include the ViT-5, as it is evaluated as part of RTLViT. 

PyTorch model compilation in reduce-overhead mode further accelerates inference on the workstation, reducing RTLViT's inference time from \SI{6.7}{\milli\second} to \SI{1.8}{\milli\second}. The U-Net improves from \SI{3.5}{\milli\second} to \SI{3.1}{\milli\second}, making the compiled RTLViT faster than the compiled U-Net. Compilation does not accelerate all architectures, as ConvNeXt's inference time increases from \SI{61.7}{\milli\second} to \SI{79.8}{\milli\second}.

We report floating-point operations (GFLOPs) per forward pass, learnable parameter counts, and model sizes in Table~\ref{tab:comparison_compute}. The three attention-based architectures compatible with real-time inference use fewer operations, fewer parameters, and less model storage than the U-Net. ViT-5 adds only 0.02 GFLOPs relative to ViT, while RTLViT requires 8.68 GFLOPs, of which the majority of operations are from the Fast AR-CNN refinement network. Compared to RTLViT, the U-Net uses $4.4\times$ more operations and $10.8\times$ more parameters. RTLViT has a \SI{5.36}{\mega\byte} model size, which is $8.8\times$ smaller than the \SI{47.12}{\mega\byte} U-Net. Model sizes are calculated for 32-bit floating-point parameters and account for all parameters and buffers, including the cached RoPE tables in ViT-5 and RTLViT. The iPhone application has a total size of \SI{6.5}{\mega\byte}, which includes the RTLViT model and one measurement for reconstruction.

\begin{table}[hbtp]
\centering
\caption{Model inference times (\SI{}{\milli\second}) across computing platforms}
\label{tab:comparison_inference}
\begin{tabular}{lcccc}
\toprule
\textbf{Model}
& \makecell{\textbf{Workstation}\\\textbf{(NVIDIA A6000)}}
& \makecell{\textbf{Compiled}\\\textbf{(NVIDIA A6000)}}
& \makecell{\textbf{Laptop}\\\textbf{(MacBook Pro M1)}}
& \makecell{\textbf{Smartphone}\\\textbf{(iPhone 17 Pro)}} \\
\midrule
ViT         
& 2.5 & 1.1 & 4.3 & 4.6 \\ 
ViT-5 
& 5.9 & 1.3 & 8.3 & N/A \\
U-Net   
& 3.5 & 3.1  & 16.2 & 16.3 \\
RTLViT (Ours)        
&  6.7 & 1.8  & 15.6 & 30.3 \\
ConvNeXt 
& 61.7 & 79.8 & 535 & N/A \\

\bottomrule
\end{tabular}
\end{table}

\begin{table}[hbtp]
\centering
\caption{Floating-point operations (GFLOPs), learnable parameter counts, and model sizes for real-time compatible architectures}
\label{tab:comparison_compute}
\begin{tabular}{lcccc}
\toprule
\textbf{Model}
& \textbf{GFLOPs}
& \textbf{Parameters} 
& \textbf{Model size (MB)}
\\ 
\midrule
ViT%
& 2.51  & 1,006,867 & 4.03 \\ %
ViT-5 
& 2.53 & 1,005,715 & 5.01 \\ %
U-Net   
& 38.16 & 11,772,307 & 47.12 \\ %
RTLViT (Ours)  
& 8.68 & 1,091,290 & 5.36 \\ %

\bottomrule
\end{tabular}
\end{table}

\clearpage

\end{document}